\documentclass[twocolumn,amsmath,amssymb]{revtex4}

\usepackage{graphicx}% Include figure files
\usepackage{dcolumn}% Align table columns on decimal point
\usepackage{bm}% bold math
\newcommand{\be}{\begin{equation}}
\newcommand{\ee}{\end{equation}}

\begin{document}
\title{Codimension Two Compactifications and the Cosmological Constant Problem}
\author{Ignacio Navarro}
\altaffiliation[E-mail address:]{ignacio.navarro@durham.ac.uk}
\affiliation{IPPP, University of Durham, DH1 3LE Durham, UK}

%\receipt{27 June 2002}
\begin{abstract}

We consider solutions of six dimensional Einstein equations with
two compact dimensions. It is shown that one can introduce
3-branes in this background in such a way that the effective four
dimensional cosmological constant is completely independent of the
brane tensions. These tensions are completely arbitrary, without
requiring any fine tuning. We must, however, fine tune bulk
parameters in order to obtain a sufficiently small value for the
observable cosmological constant. We comment in the effective four
dimensional description of this effect at energies below the
compactification scale.

\vspace{0.3cm} \noindent PACS:11.25.Mj, 98.80.Jk \hspace{6.4cm}
IPPP/03/07, DCPT/03/14.
\end{abstract}


\maketitle

{\bf 1. Introduction.} Extra dimensional models have proven to be
very fruitful in providing new ways of attacking old problems.
Despite this, some of them, like the cosmological constant problem
(CCP), still lack a compelling solution. In this paper we report
in what we believe could be some progress in obtaining an eventual
solution to this formidable puzzle.

It has been known for some time \cite{sundrum,rosskogan,lutis}
that if one introduces a 3-brane in a codimension 2 bulk around
which the background solution is rotationally symmetric, the only
effect of a non-zero brane tension will be to induce a conical
singularity (or a deficit angle) in the transverse space, and the
4 dimensional effective cosmological constant will be independent
of this tension. This scenario realizes the idea of
\cite{arkanisetc}: the curvature associated with the energy
carried by the brane is measurable by "bulk observers" only, being
spent in curving the extra dimensional manifold, and not the brane
worldvolume. However, this scenario has never been realized
satisfactorily in the literature. In \cite{sundrum} a fine tuning
relation between different brane tensions had to be imposed. In
\cite{lutis} the hope was find solutions that localized gravity in
a 3-brane in a rotationally symmetric 6D bulk, but the solutions
found always involved naked spacetime singularities or did not
compactify the extra space at all (see, however, \cite{Dvalis} for
a proposal along this lines). Other scenarios considered in the
literature for localizing gravity in a 3-brane in a 6D bulk that
are free of singularities \cite{Ruth} are not insensitive to the
deficit angle, so some fine tuning between brane and bulk
parameters has to be imposed.

In this letter we consider solutions of 6 dimensional gravity with
two compact dimensions, along the lines of the spontaneous
compactification models of \cite{weinberg,salam}. One of them is
periodic and the background is rotationally symmetric around it,
so we are able to realize the aforementioned scenario, recovering
4D gravity at low energies and in a setup free of singularities
(except for the conical ones). As has been previously discussed
\cite{lutis}, this is not a complete solution to the CCP, since we
still need to fine tune bulk parameters to obtain a small enough
value for the 4D cosmological constant. However, it is progress
because one could hope that the required fine tunings, not
involving brane parameters, are consequence of some unbroken
symmetry in the bulk ($e.g.$ supersymmetry). The outline of the
paper is as follows: in the next section we present the solution,
explain the induced conical singularity and comment on possible
bulk topologies. In section 3 we comment on the effective low
energy 4D description, and we will try to get some insight in the
effective 4D mechanism that cancels the vacuum energy coming from
"brane fields". Finally we write the conclusions.

\vspace{0.3cm}{\bf 2. Bulk Solution.}The metric ansatz we take is
of the factorizable form

\be ds^2 = \gamma(x)_{\mu \nu}dx^{\mu}dx^{\nu}+\kappa(z)_{ij} \;
dz^idz^j \ee

\noindent where latin indices run over the two extra dimensions
and greek indices over the conventional ones. We will consider an
energy-momentum tensor with a vacuum expectation value given by

\be
T_{MN}=-\left(\begin{array}{cc}\gamma_{\mu \nu}\Lambda_{\gamma} & \\
& \kappa_{ij}\Lambda_{\kappa}\end{array} \right).\label{emt}\ee

\noindent  Inhomogeneous forms for this tensor have been
considered previously in the literature
\cite{lutis,rosskogan,weinberg,salam,inhomo}, and while we will
not discuss here the origin of this energy-momentum tensor, we
will simply note that the inhomogeneity can be due to different
contributions to the casimir energy in the different directions
\cite{weinberg}, or to a vacuum expectation value of some p-form
field \cite{lutis,salam}. From Einstein equations one can check
that the metric admits a solution where the space is the direct
product of a four dimensional manifold (with metric $\gamma_{\mu
\nu}$) times a two dimensional one (with metric $\kappa_{ij}$),
both of constant curvature, being this curvatures

\be R(\gamma_{\mu \nu})=2\Lambda_{\kappa} \;\; , \;\;
R(\kappa_{ij}) = 2\Lambda_{\gamma} - \Lambda_{\kappa}.
\label{Rs}\ee

\noindent In this expressions and in what follows powers of the
higher dimensional Planck mass, $M_{Pl}^{(6D)}$, should be
understood where needed. Of direct phenomenological interest will
be the case in which the four dimensional manifold is de Sitter
space, $dS_4$, or Minkowski space, $E^{(1,3)}$, and the two
dimensional one is compact. Since Einstein equations do not fix
the topology of the space we have to make some assumptions about
it. For this compactification manifold we will consider two
different possibilities: a sphere, $S^2$, or the disk, $D_2\equiv
S^2/Z_2$. This second manifold is an orbifold obtained from the
the sphere by the identification of points in the northern and
southern hemispheres that are symmetric under reflections through
the equatorial plane. The equator of the sphere, being invariant
under this reflection, is an orbifold fixed curve that forms the
boundary of the manifold. For these spaces the metric will take
the form

\be ds^2 = dt^2 - e^{Ct}d\vec{{\bf x}}^2 - R^2_0 (d\theta^2 +
\beta^2 sin^2\theta \;d\varphi^2) \label{sol} \ee

\noindent with $\varphi$ taking values in $[0,2\pi)$ and $\theta$
ranging from $0$ to $\pi$ in the case of a sphere or from $0$ to
$\pi/2$ for the case of the disk. The constants $C$ and $R_0$ can
be determined as

\be C^2 = -{2\over 3} \Lambda_{\kappa} \;\; , \;\;
R^{-2}_0={1\over 2}\Lambda_{\kappa}-\Lambda_{\gamma}.\label{ctes}
\ee

It is clear that if we want the parameter $C$ to be small enough
to agree with observations \cite{supernova} we have to fine tune
$\Lambda_{\kappa}$ to be very small. From the previous equations
it is apparent that in principle we could choose $\Lambda_{\kappa}
= \Lambda_{\gamma}$, so the required energy-momentum tensor would
be derived from a conventional 6D cosmological constant. But in
this case, if we fix $\Lambda_{\kappa}$ to obtain a value of $C$
compatible with observations \cite{supernova} we would get a size
for the compactification manifold that is too large. In this
scenario we have to assume that $|\Lambda_{\gamma}|\gg
|\Lambda_{\kappa}|$ in order to stabilize the size of the extra
dimensions to a small enough value. It is interesting to note that
for some values of $\Lambda_{\gamma}$ one could find a solution
for the gauge hierarchy problem in the form of \cite{ADD}. This
problem would be rephrased here in finding a theory that gives
naturally values of $\Lambda_{\gamma}$ in the appropriate range.

Note also that we included an arbitrary constant $\beta$ in the
($\varphi, \varphi$) component of the metric. For any value of
this constant different from one there will be a conical
singularity at $\theta=0$ (and in $\theta=\pi$ for the case of the
sphere). The metric for the two dimensional compact manifold will
take the usual form with $\beta=1$ if we redefine $\varphi$ so it
ranges from $0$ to $2\pi\beta$, so we can think of this manifold
as being a disk (or sphere) with a "wedge" cut out. This deficit
angle gives a $\delta$-like contribution to the curvature tensor
localized at the points with $sin(\theta) = 0$. This $\delta$-like
curvature singularity will be cancelled if we introduce a 3-brane
at this positions with tension $T_0=2\pi(1-\beta)$, $i.e.$, if we
add a piece to the Lagrangian given by

\be \Delta {\cal L} = {T_0\over 2\pi \sqrt{\kappa}}\delta
\left(sin(\theta)\right) \label{brane}\ee

\noindent where $\sqrt{\kappa}$ is the volume element in the
compactification manifold (see \cite{ale} for a discussion of
delta functions in curved manifolds). We can see that the
contribution to the energy-momentum tensor of a piece in the
Lagrangian given by eq.(\ref{brane}) will be cancelled by the
contribution to the Einstein tensor induced by the deficit angle
in a number of different ways. One can compute directly the
curvature tensor for the metric (\ref{sol}) and check that a term
proportional to $(1-\beta)\delta \left(sin(\theta)\right)$ is
present in a way analogous to \cite{rosskogan}. For the sake of
completeness, and to show that this singularity can be
consistently smoothed out, here we will consider a regularization
of a 3-brane placed in $\theta=0$ and we will see that in the
infinitely thin limit, the only effect of this brane will be to
induce this conical singularity. We will not have to assume any
particular regularization procedure, since in the thin limit the
result is independent of it. We proceed as follows: we add a
finite term ($\Delta T_{M N}(\theta,\epsilon)$) to the
energy-momentum tensor such that for $\theta
> \epsilon\Rightarrow\Delta T_{M N}=0$ and

\be lim_{\epsilon \rightarrow 0} \Delta T_{M N}(\theta,\epsilon) =
-\left(\begin{array}{cc}\gamma_{\mu \nu}{T_0\over 2\pi \sqrt{\kappa}}\delta(\theta) & \\
& 0 \end{array}\right) \label{limit},\ee

\noindent that is, a regularization of the brane with width $<
\epsilon$. For $\theta > \epsilon$ the solution is given by
eqs.(\ref{sol},\ref{ctes}), while for $\theta<\epsilon$ it depends
on the regularization used. Considering an ansatz for the metric
like

\be ds^2 = dt^2 - e^{Ct}d\vec{{\bf x}}^2 - R^2_0 (d\theta^2 +
\rho(\theta)^2 \;d\varphi^2) \label{ants}\ee

\noindent  the ($0,0$) component of Einstein equations is

\be \rho ''(\theta )-\frac{R_0^2\,\left( 3\,C^2 + 4\,\Lambda
\right) \,
     \rho (\theta )}{4} = -R_0^2 \;\rho(\theta) \Delta T_{00}.  \ee

\noindent It is always possible to consider regularizations of the
brane such that one finds solutions with the ansatz (\ref{ants}),
and it is a straightforward exercise to construct a particular
regularization and solve for the function $\rho(\theta)$. We
integrate now the previous equation from $0$ to $\epsilon$, and
take the limit $\epsilon \rightarrow 0$. For doing this we only
need to take into account the following facts: $\rho(0)=0$
($\theta=0$ corresponds to a single point in the transverse
space), $\rho '(0)=1$ (the geometry is regular at the origin as it
should in any regularization of the $\delta$-like brane), for
$\theta>\epsilon$ the solution is given by (\ref{sol}) and in the
infinitely thin limit the extra contribution to the
energy-momentum tensor is given by eq.(\ref{limit}). Then one
obtains

\be 1-\beta={T_0\over 2\pi}.\label{defang} \ee

For the rest of the $(\mu,\nu)$ equations we obtain the same
result, independently of the regularization used, while the
$(\theta,\theta)$ and $(\varphi,\varphi)$ components of the
equations give a zero contribution to the matching condition in
the thin limit, since $\rho ''(\theta )$ do not appear in them,
and this is the only term that contributes.

For the case in which the compactification manifold is a sphere,
the same treatment applies to the singularity at $\theta = \pi$,
so we have to consider configurations with two "twin branes" at
the north and south poles, with equal tensions. One could argue
that the fine tuning of the two brane tensions against each other
could be regarded as natural in such symmetric configurations, but
one can also consider the compactification in the disk, so no
second 3-brane is needed. In this case one has to look at the
matching conditions for Einstein equations at the orbifold fixed
points, those with $\theta=\pi/2$. If there was a jump in the
first derivatives of the metric components with respect to the
$\theta$ coordinate at these points, we would need to assume the
presence of a 4-brane with non-zero energy-momentum tensor in
$\theta=\pi/2$. This is a subtle issue, since this could
reintroduce a fine tuning between bulk and brane parameters if the
energy carried by this 4-brane depends in the deficit angle
\cite{lutis}. However, it is straightforward to see from
(\ref{sol}) that the first derivatives of the metric components
are zero at the orbifold fixed points, and the matching conditions
are satisfied trivially, without having to consider the presence
of a 4-brane carrying any energy at this position.

\vspace{0.3cm} {\bf 3. 4D Effective Theory and the Cosmological
Constant.} From the six dimensional point of view, it is clear
that the inflation rate in the brane worldvolume only depends on
$\Lambda_{\kappa}$, and not on the brane tension, since the
parameter $C$ in the solution is fixed by eq.(\ref{ctes}), and it
is completely independent of $T_0$. The effect of the three brane
at $\theta=0$ with nonzero tension can only be to induce a conical
singularity, since it gives a contribution to the 6D Einstein
equations proportional to $\delta(\theta)$, just like a deficit
angle in the transverse space. However, from the 4D point of view,
it is not transparent how this mechanism works. Consider the low
energy theory at energies below the compactification scale. We
will have a massless 4D graviton. The equation of motion for this
mode can be extracted from the metric

\be ds^2 = \tilde{g}_{\mu \nu}(x)dx^\mu dx^\nu - R^2_0 (d\theta^2
+ \beta^2 sin^2\theta \;d\varphi^2) \label{fluc} \ee

\noindent for which the 6D curvature scalar can be expressed as

\be R^{(6D)} = R^{(4D)}(\tilde{g}) - {2\over R_0^2} -
(1-\beta){\delta(\theta)\over \sqrt{\kappa}}.\label{6DR}\ee

\noindent It is then straightforward to extract the equation for
the 4D graviton

\be R_{\mu \nu} - {1\over 2}\tilde{g}_{\mu \nu}R = -\tilde{g}_{\mu
\nu}\left({1\over 2}\Lambda_{\kappa} + {\cal A}_2^{-1}\left(T_0 -
2\pi (1-\beta)\right)\right) \label{4Dgrav},\ee

\noindent where ${\cal A}_2$ is the area of the compactification
manifold. We know from eq.(\ref{defang}) that the deficit angle
$(1-\beta)$ exactly cancels the brane tension. This, however, was
extracted from the matching conditions in the full 6D theory.
Looking at eqs.(\ref{brane},\ref{6DR}), that are intrinsically six
dimensional, it is clear why this happens, since these terms are
the only ones with a $\delta(\theta)$. But from the point of view
of the 4D low energy graviton it is a mystery why there is a
contribution to the vacuum energy, coming from the curvature of
some internal manifold, that exactly cancels the contribution
coming from fields localized in the brane, and not the
contribution coming from those living in the bulk. It is obvious
from the equation above that the 4D graviton makes no distinction
between this different contributions. We will not attempt here to
describe completely the cancellation mechanism for the "brane
part" of the vacuum energy in the low energy 4D theory, but we
will give an argument why this must happen in order to obtain a
sensible theory. If the conical singularity contribution to the
effective 4D vacuum energy did not cancel the contribution coming
from the brane, we know that the 6D Einstein equations would not
be satisfied, and this must have a reflection in the effective 4D
theory.

Remember that the bulk solution is rotationally symmetric around
the 3-brane, and this fact is crucial for the mechanism to work.
It allows us to choose the deficit angle in such a way that it
cancels the contribution to the 6D Einstein equations coming from
the 3-brane, without reintroducing any fine tuning between bulk
and brane parameters. So let's consider now the metric
(\ref{fluc}) with the replacement

\be d\varphi \rightarrow d\varphi + A_{\mu}(x)dx^{\mu}.\ee

\noindent In this way the metric will be invariant by construction
under simultaneous transformations

\be \varphi \rightarrow \varphi + \alpha(x)\ee \be A_{\mu}(x)
\rightarrow A_{\mu}(x) - \partial_\mu \alpha(x) \ee

\noindent If the energy-momentum tensor is also invariant under
these transformations, after compactification we retain this
$U(1)$ gauge symmetry for the field $A_\mu(x)$, and we can expect
a massless graviphoton in the spectrum. In this case the
graviphoton will also be present in the low energy theory and we
can extract the equation of motion for this field from the
$(\mu,\varphi)$ component of the 6D Einstein equations. At first
order in the fluctuations $A_\mu(x)$ it is

\be \nabla_\mu F^{\mu \nu}(x) =
\left(R(\tilde{g})-2\Lambda_{\kappa}\right)A^\nu(x)
\label{graviphoton}, \ee

\noindent where derivatives are covariant with respect to
$\tilde{g}_{\mu \nu}$. The previous equation shows that the
graviphoton $can$ distinguish between contributions to the vacuum
energy coming from the brane and the bulk, since it only couples
to the latter ones. If it is to be massless, the righthand side of
eq.(\ref{graviphoton}) must vanish. This condition is nothing but
eq.(\ref{Rs}), that was obtained in the full 6D theory, obtained
from a 4D point of view. We see that the requirement of the
graviphoton to be massless forces the deficit angle in
eq.(\ref{4Dgrav}) to cancel exactly the brane tension. As we said,
if this was not the case, full 6D Einstein equations would not be
satisfied, and we can expect some pathologic behaviour in the 4D
effective theory, for instance, from (\ref{graviphoton}) we see
that the graviphoton could acquire a negative mass squared.

\vspace{0.3cm}{\bf 4.Conclusions.} We have presented solutions of
six dimensional gravity with two compact dimensions. Allowing for
an inhomogeneous form for the energy-momentum tensor,
eq.(\ref{emt}), we found that the 4D inflation rate and the size
of the compact dimensions are independent parameters. Some
components of this energy-momentum tensor have to be very small to
obtain a Hubble expansion that does not contradict observations.
We did not assume any particular origin for the required
energy-momentum tensor, but there are examples in the literature
of theories were a inhomogeneous form for this tensor is generated
\cite{lutis,weinberg,salam}.

In this background it is possible to consider the presence of
$\delta$-like brane sources in such a way that the bulk solution
is rotationally symmetric around them. This fact allows the
existence of an arbitrary deficit angle in the transverse space,
that can be chosen to cancel exactly the contribution of this
sources to the 6D Einstein equations, without producing any other
effect. In this way the solution is able to adjust freely this
parameter to cancel the contribution to the 4D vacuum energy
coming from the brane tension. This is significant progress in the
longstanding CCP, since one can think of Standard Model fields
being localized on the brane and not contributing to the effective
vacuum energy. Still, one has to assume that some of the
components of the six dimensional energy-momentum tensor are
extremely small but the hope is that a model can be constructed
where this is the consequence of some (almost) unbroken
(super)symmetry in the bulk. It is remarkable also that the gauge
hierarchy problem can be related in this context to the parameter
$\Lambda_{\gamma}$ that appears in the bulk energy-momentum
tensor.

While in the full 6D theory it is clear how this mechanism works,
it is not easy to imagine how can this be seen from a 4D
perspective, since the four dimensional graviton makes no
distinction between contributions to the effective vacuum energy
coming from the brane or from the bulk. However, we have seen that
there are fields in the 4D effective theory (the graviphoton) that
$do$ make a distinction between this different contributions to
the vacuum energy. While we have not shown explicitly the
cancellation mechanism from a 4D perspective we have seen that
some pathologic behaviour can occur in the low energy effective
theory if the brane tension was to contribute to the vacuum
energy.

{\bf Acknowledgments.} I thank Ruth Gregory, Ian Kogan, Jose
Santiago and Veronica Sanz for many useful discussions and
comments.

{\bf Note Added.} While this paper was being prepared for
publication the preprint \cite{carroll} appeared, where similar
solutions to the ones presented here have been considered.

\end{document}